\input{aipcheck.tex}

\documentclass[final]
  {aipproc}

\layoutstyle{8x11double}

\SetInternalRegister\hbadness{8000} 

\newcommand\doingARLO[2][]{%
  \ifx\mmref\undefined #1\else #2\fi
}

\begin{document}

\title 
      [Constraints on the Bulk Lorentz Factor of GRB 990123]
      {Constraints on the Bulk Lorentz Factor of GRB 990123}


\author{Alicia M. Soderberg}{
  address={Institute of Astronomy, Madingley Road, Cambridge CB3 0HA, ENGLAND},
  email={ams@ast.cam.ac.uk},
}

\iftrue
\author{Enrico Ramirez-Ruiz}{
  address={Institute of Astronomy, Madingley Road, Cambridge CB3 0HA, ENGLAND},
  email={enrico@ast.cam.ac.uk},
}

\fi

\copyrightyear  {2001}

\begin{abstract}
GRB 990123 was a long, complex gamma-ray burst accompanied by an
extremely bright optical flash.  We present the collective constraints
on the bulk Lorentz factor for this burst based on 
estimates from burst kinematics,
synchrotron spectral decay, prompt radio flash observations, 
and prompt emission pulse width. 
Combination of these constraints leads to an
average bulk Lorentz factor for GRB 990123 of $\Gamma_0=1000 \pm 100$
which implies a baryon loading
of $M_{jet}=8^{+17}_{-2} \times 10^{-8}\rm M_{\odot}$. 
We find these constraints 
to be consistent
with the speculation that the optical light is emission from the
reverse shock component of the external shock.
In addition, we find the implied value of $M_{jet}$ to be in accordance with 
theoretical estimates: the baryonic loading is sufficiently small to 
allow acceleration of the outflow to $\Gamma > 100$.

\end{abstract}

\date{\today}

\maketitle

\section{Introduction}

The discovery of a prompt and extremely bright optical flash 
in GRB 990123 (Akerlof et al. 1999), implying an apparent peak (isotropic)
optical luminosity of $5 \times 10^{49}$ erg
s$^{-1}$, has lead to widespread speculation that the observed 
radiation arose from the reverse shock component of the burst. 
The reverse shock propagates into the adiabatically cooled
particles of the coasting ejecta, decelerating the shell particles and
shocking the shell material with an amount of internal energy comparable to 
that of the material shocked by the forward shock. The typical
temperature in the reverse-shocked 
fluid is, however, considerably lower than that of the forward-shocked  
fluid. Consequently, the typical frequency of the synchrotron emission
from the reverse shock peaks at lower energy. It
is believed to account for the bright prompt optical emission from GRB
990123. The reverse shock emission stops once the entire shell
has been shocked and the reverse shock reaches the inner edge of
the fluid.  Unlike the continuous forward shock, the 
hydrodynamic evolution of the reverse-shocked ejecta is more
fragile. As we will demonstrate, the temperature of the
reverse-shocked fluid is expected to be mildly-relativistic for GRB 990123 
and thus the evolution of the ejecta deviates from the Blandford \& McKee (BM) solution that determines the late profile of the decelerating shell
and the external medium.  In general, the determination of $\Gamma_0$ for optical flashes, would play an important role in discriminating between {\it cold} and
{\it hot} shell evolution. Moreover, the strong dependence of the peak time of
the optical flash on the bulk Lorentz
factor $\Gamma$ provides a way to estimate this elusive parameter. We
discuss how different constraints on $\Gamma_0$ for GRB 990123 
are consistent with optical emission from the reverse shock. 
We assume  $H_0 =65\,\, {\rm km} \, {\rm s}^{-1} \, {\rm Mpc}^{-1}$, 
$\Omega_{\rm M}=0.3$, and $\Omega_{\Lambda}=0.7$.  

\section{The Role of $\Gamma$}
Relativistic source expansion 
plays a crucial role in virtually all current
GRB models (Piran 1999; M\'esz\'aros 2001). 
The Lorentz factor is not, however,  well determined
by observations. The lack of apparent photon-photon attenuation up to 
$\approx 0.1$ GeV implies only a lower limit $\Gamma \approx 30$
(M\'esz\'aros, Laguna \& Rees 1993), while the observed pulse width
evolution in the gamma-ray phase eliminates scenarios in which  
$\Gamma >> 10^{3}$ (Lazzati, Ghisellini \& Celotti 2001; Ramirez-Ruiz
\& Fenimore 2000). The initial Lorentz factor is set by the baryon
loading, that is $m_0 c^2$, where $m_0$ is the mass of the
expanding ejecta. This energy
must be converted to radiation in an optically-thin region, as the
observed bursts are non-thermal.  The radius of transparency of the
ejecta is 
\begin{equation}
R_{\tau}= \left({\sigma_T E \over 4 \pi m_p c^2 \Gamma_0}
\right)^{1/2}  \approx 10^{12}-10^{13} {\rm cm}, 
\end{equation}
where $E$ is the
isotropic equivalent energy generated by the central site. The highly
variable $\gamma$-ray light curves can be understood in terms of
internal shocks produced by 
velocity variations within the relativistic outflow 
(Rees \& M\'esz\'aros 1994).  
In an unsteady outflow, if $\Gamma$
were to vary by a factor of $\approx$ 2 on a timescale $\delta T$,
then internal shocks would develop at a distance $R_{\rm i} \approx
\Gamma^2 c \delta T \ge R_\tau$. 
This is followed by the development of a
blast wave expanding into the external medium, and a
reverse shock moving back into the ejecta.   
The inertia of the swept-up external matter decelerates the shell
ejecta significantly by the time it reaches the  
deceleration radius (M\'esz\'aros \& Rees 1993),
\begin{equation}
R_{\rm d}=\left( {E \over n_0 m_{p}c^2\Gamma_0^2}\right)^{1/3} \approx
10^{16}-10^{17} {\rm cm}. 
\end{equation}
Given a certain external baryon density $n_0$, the initial Lorentz
factor then strongly determines where both
internal and 
external shocks develop. Changes in $\Gamma_0$ will modify the
dynamics of the shock deceleration and the manifestations of the
afterglow emission.

\section{Reverse Shock emission and $\Gamma _0$}

It has been predicted that the reverse shock produces a  
prompt optical flash brighter than 15th magnitude with
reasonable energy requirements of no more than a few $10^{53}$ erg 
emitted isotropically (M\'esz\'aros \& Rees 1999; Sari \& Piran 1999a;
hereinafter MR99 and  SP99).
The forward shock emission is continuous, but
the reverse shock terminates once the shock
has crossed the shell and the cooling frequency has dropped below the observed
range. The reverse shock contains, at the time it crosses the
shell, an amount of energy comparable to that in the 
forward one. However,
its effective temperature is significantly lower (typically by a
factor of $\Gamma$). Using the shock jump
conditions and assuming that the electrons and the magnetic field acquire a
fraction of the
equipartition energy $\varepsilon_e$ and $\varepsilon_B$ respectively, 
one can
describe the hydrodynamic and magnetic conditions behind the shock.

The reverse shock synchrotron 
spectrum is determined by the 
ordering of three break frequencies, the self-absorption 
frequency $\nu_a$, the cooling frequency
$\nu_c$ and the characteristic synchrotron frequency
$\nu_m$, which are easily calculated by comparing them to
those of the forward shock (MR99; SP99; Panaitescu \& Kumar 2000). 
The equality of energy density across the contact
discontinuity suggests that the magnetic fields in both regions are
of comparable strength. 

Assuming that the
forward and reverse shocks both move with a similar Lorentz
factor, the reverse shock  synchrotron frequency is given by
\begin{equation}
v_m=5.840\times 10^{13} \epsilon_{e,-1}^2 \epsilon_{B,-2}^{1/2}
n_{0,0}^{1/2} \Gamma_{0,2}^2 (1+z)^{-1/4} {\rm Hz}, 
\end{equation}
\noindent
while the cooling frequency $\nu_c$ is equal to that of the
forward shock.  Here we adopt the convention $Q
= 10^x\,Q_x$ for expressing the physical parameters, using cgs units.  
The spectral power
$F_{\nu_m}$ at the  characteristic synchrotron frequency is 
\begin{equation}
F_{\nu_m}=4.17 D_{28}^{-2} \epsilon_{B,-2}^{1/2} E_{53} n_{0,0}^{1/2}
\Gamma_{0,2} (1+z)^{3/8} {\rm Jy}. 
\end{equation} 
The distribution of the injected electrons is assumed to be
a power law of index $-p$, above a minimum Lorentz factor
$\gamma_i$.
For an adiabatic blast wave, the
corresponding spectral flux at a given 
frequency above $\nu_m$ is $F_\nu \approx F_{\nu_m}(\nu / \nu_m
)^{-(p-1)/2}$, while below  $\nu_m$ is characterised by a
synchrotron tail  with  $F_\nu \approx F_{\nu_m}(\nu /
\nu_m)^{1/3}$. Similar relations to those found for a radiative forward shock
hold for the reverse shock (Kobayashi 2000; hereinafter K00).

Unlike the synchrotron 
spectrum, the afterglow light curve at a fixed frequency strongly
depends on the hydrodynamics of the relativistic shell, which determines the
temporal evolution of the break frequencies $\nu_m$ and $\nu_c$. The 
forward shock is always highly relativistic  and thus is successfully
described using the relativistic generalisation of theory of 
supernova remnants. In contrast, the reverse shock can be mildly
relativistic.  In this regime, the  shocked shell is unable to heat
the ejecta to  sufficiently high temperatures and its evolution
deviates from the BM solution (Kobayashi \& Sari 2000; hereinafter
KS00). Shells  satisfying 
\begin{equation}
\xi \approx 0.01
E_{52}^{1/6} \Delta_{11}^{-1/2}\Gamma_{0,2}^{-4/3} n_{0,1}^{-1/6}   > 1,
\end{equation}
are likely to have a Newtonian reverse shock\footnote{ It should be
remarked that equations (3) and (4) refer to this reverse shock
regime. Equations for the relativistic case are 
relatively similar, the biggest discrepancy being that the peak flux
is inversely proportional to $\Gamma$ (see equations (7)-(9) of
K00).}, otherwise the reverse shock is relativistic and it
considerably decelerates the ejecta. The width of the shell, $\Delta$,
can be inferred directly from the observed burst duration by
$\Delta=cT_{\rm dur}/(1+z)$ assuming the shell does not undergo
significant spreading (Piran 1999). 

If $\xi > 1$, then the reverse
shock is in the sub-relativistic temperature regime
for which there are no known analytical solutions.  
In order to constrain the evolution of $\Gamma$ in this regime it is
common to assume $\Gamma \propto R^{-g}$ where $3/2 \le g \le 7/2$
(MR99; KS00). For an adiabatic
expansion, $\Gamma \propto T^{-g/(1+2g)}$ and so $\nu_m \propto
T^{-3(8+5g)/7(1+2g)}$ and $F_{\nu_{m}} \propto
T^{-(12+11g)/7(1+2g)}$. 
The spectral flux at a given frequency
expected from the reverse shock gas drops then as
$T^{-2(2+3g)/7(1+2g)}$ below $\nu_m$ and $T^{-(7+24p+15pg)/14(1+2g)}$
above. For a typical spectral index $p=2.5$, the flux decay
index varies in a relatively narrow range ($\approx 0.4$) between
limiting values of $g$.
\section{Constraints on the $\Gamma_0$ of GRB 990123}
Despite ongoing observational attempts, the optical flash associated
with GRB 990123 remains the only event of its kind detected to date. 
Observations of this optical flash appear to be in good agreement
with early predictions for the reverse shock emission (Sari \& Piran
1999b).   
As a result, numerous studies have been done on this event in which  
reverse shock theory has been applied to burst observations in order
to constrain physical parameters and burst properties, including
$\Gamma_0$.  Current estimates on the bulk Lorentz factor for GRB
990123 stretch over nearly an order of magnitude, with values ranging
from $\approx 200$ (SP99) to $\approx 1200$
(Wang, Dai \& Lu 2000).  It should be noted, however, that these estimates
were made before accurate burst parameters for GRB 990123 were known, 
and consequently they include approximations and parameters
from other GRB afterglows. 
\begin{figure}
   \includegraphics[width=0.44\textwidth]{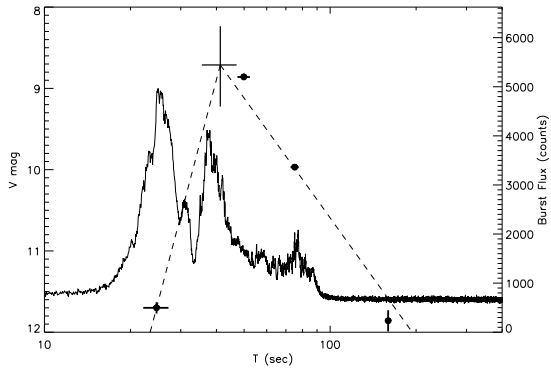}
   \caption{BATSE and ROTSE light curves for GRB 990123 as a function of
time from the BATSE trigger. Dashed lines represent theoretical
predictions for the rise  $\propto T^{3p-3/2}$ and decay $\propto 
T^{-(21+73p)/96}$ of an adiabatic reverse shock light curve, assuming
the shell is thin and cooling slowly.  We predict that the optical flash
peaked $\approx 41\pm 6$ seconds after the trigger. }
\end{figure}

By fitting multi-frequency afterglow light curves, physical parameters
for 8 GRBs have recently been reported (Panaitescu $\&$
Kumar 2001b, hereinafter PK01). Best fit values presented for GRB 990123 
are $E_{j,50}=1.5^{+3.3}_{-0.4}$ (initial jet energy),
$\theta_0=2.1^{+0.1}_{-0.9}$, $n_{0,-3}=1.9^{+0.5}_{-1.5}$ ,
$\epsilon_{e,-2}=13^{+1}_{-4}$,
$\epsilon_{B,-4}=7.4^{+23}_{-5.9}$, and
$p=2.28^{+0.05}_{-0.03}$ with a 
rough estimate for the bulk Lorentz factor of $\Gamma_0=1400 \pm 700$ (Panaitescu $\&$ Kumar 2001a).  
Using these physical parameters, we present a comprehensive examination
of the constraints on $\Gamma_0$ and report a best-fit value based on
an analysis of these constraints.   

\paragraph{Time of Peak Flux}
Observational estimates for the time of peak flux enabled a
measurement of the initial Lorentz factor with reasonable accuracy
using the physical
parameters specific to GRB 990123.
Assuming the optical flash was the result of the
reverse shock, the initial bulk Lorentz factor 
\begin{equation}
\Gamma_0=237~E_{52}^{1/8}~n_{0,0}^{-1/8}~T_{{\rm d},1}^{-3/8}~(1+z)^{3/8}
\end{equation}
where $T_{\rm d}$ is the time of peak flux in the observer frame.
Light curves for the optical flash and $\gamma$-ray emission are shown
in Figure 1. Dashed lines represent theoretical predictions for the
rise  $\propto T^{3p-3/2}$ and decay $\propto 
T^{-(21+73p)/96}$ of an adiabatic reverse shock light curve, assuming
the shell is thin, i.e. $\Delta < (E/(2 n_0 m_{\rm p} c^2
\Gamma_0^{8}))^{1/3}$, and cooling slowly (K00). Using the
recently reported physical parameters, one finds  GRB 990123 to have 
a marginal thickness, as predicted by K00. The
observed rise time is, however, in good agreement with that of a thin
shell $\approx T^{5.5}$ (in contrast
with $\approx T^{1/2}$ for a thick shell). 
The shape of the light curve is determined
by the time evolution of the three spectral break
frequencies, which in turn depend on the hydrodynamical evolution of
the fireball. In the case of GRB 990123, the typical synchrotron
frequency $\nu_m=1.5\times 10^{14}$ Hz is well below the cooling 
frequency $\nu_c= 1.0\times 10^{19}$
Hz, and therefore places the burst in a regime with a flux decay
governed by the relation $F_{\nu} \propto  T^{-(21+73p)/96}$. This
implies a decay of 
$\approx T^{-2}$ for the optical flash of GRB 990123.  Applying these
light curve predictions to the prompt optical data and taking observational
uncertainties as well as burst parameter uncertainties into account,
we predict that the optical flash peaked $T_{\rm peak}\approx 41\pm
6$ s after the GRB started. Substitution for $T_{\rm peak}$ in
equation (6) gives $\Gamma_0=770 \pm 50$.

\paragraph{Synchrotron Spectral Decay}
Observations of the optical peak brightness enable further accurate
constraints on  the value of $\Gamma_0$. The synchrotron spectrum
from relativistic electrons comprises four power-law
segments, separated by three critical frequencies. The prompt optical
flash in GRB 990123 is observed at a frequency that falls
well below the cooling frequency, but above the typical synchrotron
frequency:  $\nu_a < \nu_m <\nu_{\rm obs}<\nu_c$. The
synchrotron spectrum for this spectral segment 
is given by $F_{obs}=F_{\nu_{m}} (\nu_{obs}/\nu_m)^{(p-1)/2}
(1+z)^{1/2-p/8}$ where $\nu_{obs}$ is taken to be the ROTSE optical
frequency. Assuming the optical peak flux $F_{obs}$ observed in
GRB 990123 is radiation arising from the reverse shock, we find
$\Gamma_0=1800^{+600}_{-500}$. 

\paragraph{Radio Flare Observations}
In addition to the optical flash associated with GRB 991023, a short ($\sim 30$ hr.) radio
flare was also observed (Kulkarni et al., 1999).  Since the emission
did not grow stronger with time thereby demonstrating the properties
of a typical radio afterglow, it was subsequently proposed 
that both the prompt 
optical emission and the radio flare arose from the reverse shock (SP99).
As the reverse shock cools the emission shifts to lower frequencies which
rapidly approach the observational radio band. The received flux
peaks when $\nu_{\rm m}$ crosses the band while decreasing 
mildly relativistically as $\nu_m \propto
T^{-3(8+5g)/7(1+2g)}$.  
At $t < 1.2$ days after the initial GRB trigger, the radio flare from 990123
was observed to be increasing thereby implying $\nu_{\rm m} > \nu_{\rm obs}$.
Using eqn (3) and the evolution of $\nu_{\rm m}$, we compute the $\Gamma_0$ which places the peak frequency 
in the radio band at $t\sim1.2$ days.  We find $\Gamma_0=2000^{+400}_{-200}$.  
\begin{figure}
   \includegraphics[width=0.44\textwidth]{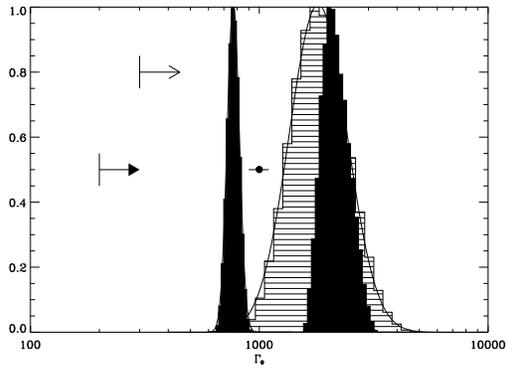}
   \caption{Collective constraints on $\Gamma_0$ for GRB 990123. These
include estimates from the burst kinematics (narrowest distribution),
synchrotron spectral decay (widest distribution), radio flash observations (medium width distribution), prompt emission pulse width (filled arrow), and jet modelling (unfilled arrow).  We find a best fit value of $\Gamma_0=1000\pm 100$.}
\end{figure}

\paragraph{Prompt Emission Pulse Width}
Although observations of a reverse shock induced optical peak enable 
fairly accurate calculations of the bulk Lorentz factor, it remains
possible to obtain information on $\Gamma_0$ in situations when an 
optical flash has not been detected.  
Consider an internal shock which produces an instantaneous burst of
isotropic $\gamma$-ray emission at a time, $t$, and radius, $R_i$, in
the frame of the central engine.  The kinematics of colliding shells 
implies that although photons are emitted simultaneously, the
curvature of the emitting shell spreads the arrival time of the
emission over a period of $\Delta T_{\rm p}$, thereby producing the 
observed width in individual pulses.  
The delay in arrival time between on-axis photons and those at $\theta
\approx \Gamma^{-1}$ is a function of the radius of emission and the
Lorentz factor according to: 
$\Delta T_{\rm p}/(1+z)=R_{\rm i}/(2c\Gamma^2)$ where $\Gamma=\Gamma_0$ in
the early phase of the expansion (Fenimore, Madras \& Nayashin 1996).  
In order to allow photons to escape, $R_{\rm i}$ must be larger than the
radius of transparency $R_{\tau}$.  This imposes a lower limit on the
initial bulk Lorentz factor such that: $\Gamma_0 >
(R_{\tau}/2c \Delta T_{\rm p})^{1/2} (1+z)^{1/2}$. 
We determine $\Delta T_{\rm p}$ for GRB 990123 by measuring the average pulse
width through autocorrelation methods based on those described in
Fenimore, Ramirez-Ruiz \& Wu (1999) and find $\Delta T_{\rm p}\approx
0.45$ s.  Using the burst parameters to estimate $R_{\tau}$ from equation (1)
and applying the inequality relation defined above, we find a lower
limit of $\Gamma_0 > 200$.   
Figure 2 displays the collective constraints on $\Gamma_0$ for GRB
990123. An additional lower limit of $\Gamma_0 >
300$ constraint from afterglow modelling is also included (Panaitescu \& Kumar 2002). The combination of these constraints leads to an
average bulk Lorentz factor for GRB 990123 of $\Gamma_0=1000 \pm 100$
which implies $R_{\tau} \approx  1.3 \times 10^{14}$ and a baryon loading
of $M_{jet}=8^{+17}_{-2} \times 10^{-8}\rm M_{\odot}$.  

\section{Conclusions}
We have shown that the collective constraints for the bulk Lorentz factor 
of GRB 991023 are compatible with current reverse shock theory.  
In addition, our best fit value of $\Gamma_0 \approx 1000$ provides
confirmation of the ultra-relativistic nature of GRBs.  The implied
values for $R_{\tau}$ and $M_{jet}$ are in accordance with 
GRB theory: the radius of transparency is within theoretical
estimates and the baryonic loading is sufficiently small to 
allow acceleration of the outflow to $\Gamma > 100$.
As we have discussed, the bulk Lorentz factor plays a crucial role 
throughout all stages of GRB evolution, and therefore the ability to constrain 
$\Gamma_0$ provides clues on the nature of gamma-ray bursts.

We thank A. Blain, D. Lazzati, M.~J. Rees and E. Rossi for helpful
conversations. AMS  was supported by the NSF GRFP. ER-R
thanks CONACYT, SEP and the ORS for support.


\end{document}